\documentclass[a4paper,10pt]{amsart}

\usepackage[norelsize]{algorithm2e}
\usepackage{graphicx}
\usepackage{geometry}
\newtheorem{Theorem}{Theorem}

\begin{document}

\title{Computing the Power Distribution in the IMF}

\author{Sascha Kurz}
\address{Sascha Kurz, Mathematisches Institut, Universit\"at Bayreuth, 95440 Bayreuth, Germany. E-mail: sascha.kurz@uni-bayreuth.de, 
Phone: +49\,921\,55\,7353, Fax: +49\,921\,55\,7353, Homepage: http://www.wm.uni-bayreuth.de/index.php?id=sk}

\begin{abstract}
The International Monetary Fund is one of the largest international 
organizations using a weighted voting system. The weights of its 
188 members are determined by a fixed amount of basic votes plus some 
extra votes for so-called Special Drawing Rights (SDR). On January 26, 
2016, the conditions for the SDRs were increased at the 14th General 
Quota Review, which drastically changed the corresponding voting weights. 
However, since the share of voting weights in general is not equal to 
the influence, of a committee member on the committees overall 
decision, so-called power indices were introduced. 
So far the power distribution of the IMF was only computed by 
either approximation procedures or smaller games than then entire Board of 
Governors consisting of 188 members. We improve existing algorithms, based 
on dynamic programming, for the computation of power indices and provide the 
exact results for the IMF Board of Governors before and after the increase of 
voting weights. Tuned low-level details of the algorithms allow the repeated 
routine with sparse computational resources and can of course be applied to other 
large voting bodies. It turned out that the Banzhaf power shares are rather sensitive 
to changes of the quota.

\medskip

\end{abstract}

\maketitle        

\section{Introduction}

The International Monetary Fund (IMF) was formed in 1944 at the Bretton Woods Conference. Currently 
this international organization consists of 188 countries as members. Its highest decision-making body, 
i.e., the \emph{Board of Governors},  makes its decisions by weighted voting. The weights are 
composed of \emph{basic votes}, which are equal for each member and sum up to 5.502 percent of the total 
votes, and one additional vote for each Special Drawing Right (SDR) of 100,000 of a member country's quota 
(the IMF term for the country's financial stake, c.f.~\cite{leech2013new}). On January 26, 2016, the 
conditions for the SDRs were increased at the Board Reform Amendment, which drastically changed the 
corresponding voting weights. In general the weight of a country can be a poor proxy for its influence 
in a weighted voting game.\footnote{Consider, e.g., a committee, where the weight shares are 49\%, 49\%, 
and 2\%. For simple majority a least two out of the three committee members are needed in order to push
through a proposal, i.e., the influences are equal contrary to the voting weights.} To this
end, so-called power indices were introduced in order to measure the \emph{influence} or \emph{power} 
of a committee member in a committee making its decisions via binary voting, i.e., each member can say 
{\lq\lq}yes{\rq\rq} or {\lq\lq}no{\rq\rq} to a given proposal. As the idea of power and influence 
is not defined unambiguously, several power indices were introduced in the literature. Arguably, the 
Shapley-Shubik and the Banzhaf index are two of the most frequently applied power indices. 
%% Besides these 
%% two measure several more power indices have been proposed in the literature so far. 
Unfortunately, the 
evaluation of such a power index is a computational hard problem in general.\footnote{To be more precise, 
the computation of the power indices treated in this paper is NP-hard in the sense of computational complexity
theory. We give a brief justification at the end of Section~\ref{sec_preliminaries}.} And indeed, we are not aware of 
any paper, where either the Shapley-Shubik or the Banzhaf index of the IMF Board of Governors has been computed 
exactly.  Approximation procedures were applied in \cite{leechms, leech2013new}. The \emph{Executive Board} was, e.g., studied 
in \cite{aleskerov2008actual}. In this paper we will compute the exact numerical values of both 
 power indices for the IMF Board of Governors corresponding to voting weights slightly after and before   
the meeting on January 26, 2016. As the quota and voting shares will change as members pay their quota
increases, see \url{https://www.imf.org/external/np/sec/memdir/members.aspx}, we list the used voting weights 
in tables~\ref{table_voting_weights_part1}-\ref{table_voting_weights_part4}.\footnote{The voting weights 
were accessed at the official website \url{https://www.imf.org/external/np/sec/memdir/members.aspx}. The numbers 
were retrieved on February 17, 2016 and on July 27, 2015, respectively.}

Algorithms for the efficient computation of power indices in voting games have been studied extensively 
in the literature. By looping over all $2^n$ subsets of players, the Shapley-Shubik index of a fixed 
player can be easily computed in $O(n\cdot 2^n)$ time. The straight-forward computation of the Banzhaf 
index of a fixed player can be performed in $O(n^2\cdot 2^n)$ time. For weighted voting games these 
computation complexities were reduced to $O\!\left(n\cdot \sqrt{2}^{\,n}\right)$ and $O\!\left(n^2\cdot \sqrt{2}^{\,n}\right)$ in 
\cite{klinz2005faster}, respectively. Assuming that all weights are integers and taking the sum of voting 
weights $C$ into account, more refined complexity results can be obtained. Several algorithms based on 
generating functions were implemented in \texttt{Mathematica}, see \cite{tannenbaum1997power}. Those algorithms 
are fast if the subsets of players attain only few different weight sums. The number of different weight sums 
is clearly upper bounded by $C+1$. If almost all possible weight sums are attained, then one can use the related 
but conceptually easier concept of dynamic programming, see \cite{matsui2000survey} for a survey.\footnote{Actually, the 
only difference between the generating function and the dynamic programming approach is that the former 
utilizes the fast-access data structures for polynomials with few coefficients implemented in computer algebra systems. The 
generating function approach dates back at least to \cite{mann1962values}, where it was applied onto the 
electoral college.} 
With this, the Shapley-Shubik index of fixed player can be determined in $O(n^2 q)$ time and $O(nq)$ space, where 
$q\le C$ denotes the \emph{quota} of a weighted voting game. The Banzhaf index of a fixed player can be computed 
in $O(nq)$ time and $O(q)$ space. In \cite{uno2012efficient} these complexity bounds are maintained for the computation 
of the respective power indices for all $n$ players. We slightly improve upon these complexity bounds by replacing 
$q$ by $\min(q,C-q+1),$\footnote{For the IMF Board of Governors we have $q=0.85\cdot C$, so that we obtain an acceleration 
of a factor of $0.85/0.15\approx 5.67$. The memory requirements are reduced by the same factor.} provide an easy to 
understand description, and extend the analysis to further power indices. For practical efficiency we go into low-level 
details of the algorithms and discuss their impact on the running time for the IMF example.

The remaining part of this paper is structured as follows. In Section~\ref{sec_preliminaries} we briefly introduce 
simple games as models for voting systems and some related notation. 
%%For more details we refer the interested reader e.g.\ to \cite{taylor1999simple}. 
After introducing the defining equations for the power indices, we 
consider algorithms for their computation in Section~\ref{sec_algorithms}. These are essentially based on counting the number 
of coalitions per weights and size by dynamic programming techniques. After stating our computational results in 
Subsection~\ref{subsec_computational_results} we draw a conclusion in Section~\ref{sec_conclusion}.
The weights of the considered voting games and the resulting power distributions are outsourced into an appendix due to their 
large size. 

\section{Preliminaries}
\label{sec_preliminaries}
Let $N=\{1,\dots,n\}$ be the set of players. A \emph{simple game} (on $N$) is a mapping $v:2^N\rightarrow\{0,1\}$ 
with $v(\emptyset)=0$, $v(N)=1$, and $v(S)\le v(T)$ for all $\emptyset\subseteq S\subseteq T\subseteq N$. A subset 
$S\subseteq N$ is called \emph{coalition} and 
%%mostly 
represents the set of {\lq\lq}yes{\rq\rq}-voters. A coalition 
$S$ is called \emph{winning} if $v(S)=1$ and \emph{losing} otherwise. 
%%If $S\subseteq N$ is winning but all proper 
%%subsets of $S$ are losing, we speak of a \emph{minimal winning coaltion}. Similarly, $T\subseteq N$ is a 
%%\emph{maximal losing coalition} if $T$ is losing and all of its proper supersets are winning. 
A simple game 
$v$ is \emph{weighted} if there exist $q,w_1,\dots,w_n\in\mathbb{R}_{\ge 0}$ such that $v(S)=1$ iff $w(S)\ge q$ 
for all $S\subseteq N$, where $w(S):=\sum_{i\in S} w_i$. The $w_i$ are called \emph{weights} (for player~$i$) and 
$q$ is called \emph{quota}. We write $v=[q;w_1,\dots,w_n]$ and remark that weights and quota are far from being unique, so that 
we speak of a representation $(q,w)$ for $v$. A representation with $q\in \mathbb{N}$, $w\in\mathbb{N}_{\ge 0}^n$ is 
called \emph{integer representation}. It is well known that each weighted game admits an integer  
representation.\footnote{Let $(q,w)$ be a representation of $v$, let $\alpha$ the maximum weight of losing 
coalition and $\beta$ the minimum weight of a winning coalition. Increase the weights by at most $(\beta-\alpha)/2n>0$ 
so that they become rational numbers. As quota chose an arbitrary rational number strictly between the 
new minimum weight of a winning coalition and the new maximum weight of a losing coalition. Multiplication with the 
common denominator yields an integer representation of $v$.} We speak of a \emph{minimum sum integer representation} 
if the sum of weights is minimized within the class of all integer representations. Those representations need 
not to be unique in general if the number of players is not too small, see e.g.~\cite{kurz2012minimum}.
If $v(S\cup\{i\})\ge v(S\cup\{j\})$ for all 
$S\subseteq N\backslash \{i,j\}$ we write $i\succeq j$, which defines a partial order.  If this ordering is complete 
we call the simple game $v$ \emph{complete} and remark that all weighted games are complete. 
A player $i\in N$ is 
called a \emph{null player} (in a simple game $v$), iff $v(S)=v(S\cup\{i\})$ for all $S\subseteq N\backslash\{i\}$. 
Two players $i,j\in n$ are called \emph{equivalent}, denoted as $i\sim j$, if $i\succeq j$ and $j\succeq i$. If each 
winning coalition contains a certain player $i$, she is called \emph{veto player}.

Next we briefly introduce the used power indices. 
%%For some more details and references to the corresponding introducing 
%%papers we refer the interested reader to e.g.~\cite{Kurz2016}.  
The \emph{Shapley-Shubik} index of player $i$ is given by
\begin{equation}
  \label{eq_ssi}
  \operatorname{SSI}_i(v)=\frac{1}{n!}\cdot \sum_{S\subseteq N\backslash\{i\}} |S|!\cdot (n-|S|-1)!\cdot(v(S\cup\{i\})-v(S)).
\end{equation}
The \emph{absolute Banzhaf index} of player $i$ is given by 
\begin{equation}
  \label{eq_absolute_bz}
  \operatorname{Bz}_i^{\text{a}}(v)=\frac{1}{2^{n-1}}\cdot \sum_{S\subseteq N\backslash\{i\}} v(S\cup\{i\})-v(S).
\end{equation}
If we call a coalition $S\subseteq N\backslash\{i\}$ an \emph{$i$-swing} if $S$ is losing and $S\cup\{i\}$ winning, then 
$\operatorname{Bz}_i(v)$ is equal to the number of $i$-swings divided by the number of coalitions with(out) player~$i$. 
Normalizing to sum $1$, we obtain the (relative) \emph{Banzhaf index} of player~$i$: 
\begin{equation}
  \label{eq_bz}
  \operatorname{Bz}_i(v)=\operatorname{Bz}_i^{\text{a}}(v) / \sum_{j=1}^n \operatorname{Bz}_j^{\text{a}}(v). 
\end{equation}
%%The performed normalization is applicable for all mappings $\mathcal{P}^{\text{a}}_i$ with $\mathcal{P}_i^{\text{a}}(v)\ge 0$ 
%%and $\sum_{j=1}^n \mathcal{P}_j^{\text{a}}(v)>0$ for all (weighted) simple games $v$ on $n$ players. We can set 
%%$\mathcal{P}_i^{\text{a}}(v)=\mathcal{P}_i^{\text{a}}(v)/\sum_{j=1}^n\mathcal{P}_j^{\text{a}}(v)$, see e.g.~\cite{Kurz2016}.

The two power indices have the property that they sum up to one and assign a value of zero to a player if and only 
if she is a null player. Since it is NP-hard to decide whether a player is a null player in a given weighted game, see 
e.g.~\cite{chalkiadakis2011computational}, the computation of the used power indices is at least NP-hard. We remark that 
the equivalent players attain the same Shapley-Shubik or Banzhaf index. 

%% The eight introduced power indices all have the property that they sum up to one and assign a value of zero to null players. The 
%% nucleolus may assign a value of zero to non-null players while all other indices do not. Since it is NP-hard to decide whether a
%% player is a null player in a given weighted game, see e.g.~\cite{chalkiadakis2011computational}, the computation of the latter power 
%% indices is at least NP-hard. For the NP-hardness for the nucleolus we refer to e.g.~\cite{elkind2009computing}. For all eight power 
%% there exist pseudo-polynomial time algorithms for weighted games. For the nucleolus we again refer to \cite{elkind2009computing} 
%% while we give examples for the other power indices in the next section. 

\section{Algorithms}
\label{sec_algorithms}

Assume that we have a weighted game $v=[q;w]$ on $n$ players in integer representation, where we set $C=\sum_{i=1}^n w_i$.
As the complexity of our subsequent algorithms will depend on $\Delta:=\min(q,C-q+1)$ it would be beneficial to have a minimum 
sum integer representation at hand. However, it is not clear if minimizing the integer representation pays off 
for the computation of power indices, c.f.~\cite{mostly_sunny}, where this is proposed as a promising strategy. So, here we 
propose to perform the following computationally cheap preprocessing steps at the very least. At first we reduce the weights that 
are larger then the quota by setting $q'=q$ and $w_i'=\min(q,w_i)$ for all $i\in N$. Next we guarantee that the weights are not 
too much larger than $C-q$. If $w_i>C-q$, then player $i$ is a vetoer and we set $w_i'=C-q+1$, $q'=q-w_i+w_i'$, and $w_j'=w_j$ 
for all $j\in N\backslash\{i\}$. Both operations can be performed in $O(n)$. The power indices used in this paper do not 
only assign zero power to all null players but are \emph{null player preserving}, i.e., if $v'$ arises from $v$ by adding 
null player $i$, then we have $\mathcal{P}_j(v')=\mathcal{P}_j(v)$ for all $j\neq i$. Nevertheless, it is NP-hard to detect 
null players we can efficiently remove players with a zero weight, so that we can assume $1\le w_i\le \Delta$ %%\min(q,C-q+1)$ 
in the following, i.e., we have $C\ge n$.

%% In some parts of the subsequent algorithms it is beneficial to assume an ordered sequence of weights, i.e., $w_1\le\dots \le w_n$ 
%% or $w_1\ge \dots\ge w_n$. By applying \emph{bucket sort} we can compute such an ordering in $O(n+\Delta)=O(\Delta)$ 
%% time and $O(n+\Delta)=O(\Delta)$ space, where we set $\min(q,C-q+1)=:\Delta$ as an abbreviation.

%% The veto players can be determined in $O(n)$ time by first computing $C$ and then by checking $C-w_i\ge q$ for all $i\in N$. 
%% Since we can easily check that the United States are the unique veto player in both $\text{IMF}_{2015}$ and $\text{IMF}_{2016}$, 
%% c.f.~tables~\ref{table_voting_weights_part1}-\ref{table_voting_weights_part3}, the nucleolus is easily determined.

In the following subsections we present the algorithmic details how to compute the power indices efficiently.

\subsection{Counting coalitions per weight}
\label{subsec_coalitions_per_weight}
Let $c(x)$ denote the number of coalitions of a given weighted game $v$ attaining weight $x$. By 
Algorithm~\ref{alg:coalitions_per_weight_forward} we can compute $c(x)$ for all $0\le x\le q$ in 
$O(nq)$ time and $O(q+n)$ space, where we assume that we have precomputed the terms $\min\{q,\sum_{j=1}^i w_j\}$ 
for all $i\in N$.
 
\begin{algorithm}[htp]
\SetAlgoNoLine
\KwIn{$q$, $w$, $n$}
\KwOut{$c(x)$ for $0\le x\le q$}
$c(0)\gets 1$\;
\For{$1\le x\le q$}{
    $c(x)\gets 0$\;
    }
\For{$i$ from $1$ to $n$}{
   \For{$x$ from $\min\{q,\sum_{j=1}^i w_j\}$ to $w_i$}{
       $c(x)\gets c(x)+c(x-w_i)$\;       
       }
   }
\caption{Forward counting of coalitions per weight}
\label{alg:coalitions_per_weight_forward}
\end{algorithm}

Similarly we can compute the respective counts starting from weight $C$, see Algorithm~\ref{alg:coalitions_per_weight_backward} 
that needs $O(n\cdot(C-q+1))$ time and $O(C-q+1+n)$ space.   

\begin{algorithm}[htp]
\SetAlgoNoLine
\KwIn{$q$, $w$, $n$}
\KwOut{$c(x)$ for $q\le x\le C$}
$c(C)\gets 1$\;
\For{$q\le x\le C-1$}{
    $c(x)\gets 0$\;
    }
\For{$i$ from $1$ to $n$}{
   \For{$x$ from $\max\{q+w_i,C-\sum_{j=1}^{i-1} w_j\}$ to $C$}{
       $c(x-w_i)\gets c(x)+c(x-w_i)$\;       
       }
   }
\caption{Backward counting of coalitions per weight}
\label{alg:coalitions_per_weight_backward}
\end{algorithm}

For the ease of notation we assume that the basic arithmetic operations for integers not too much larger than $C$ can be performed 
in $O(1)$ time and space. However, the values stored in $c(x)$ can grow very quickly, i.e., we have 
$2^n\ge \max_{0\le x\le C} c(x)\ge 2^n/(C+1)$. So, we should count $\Theta(n)$ for each addition or subtraction. To avoid 
technical complications in the exposition and in order to be comparable with the related literature we also assume that all 
basic arithmetic operations for integers can be performed in constant time. From a practical point of view we have to deal with 
the corresponding problems nevertheless. In our application of the IMF we have $n=188$, so that the values of $c(x)$ do not fit
into the  standard, simple data types on a 64-bit system. Since the overhead of a general-purpose arbitrary-precision arithmetic 
is quite large, we directly implement the most frequently used  basic operations as follows. We choose different primes $p_1,\dots, p_l$, 
such that all occurring numbers are between $0$ and $-1+\prod_{i=1}^l p_i$. During the computation we perform all basic operations modulo $p_i$ 
for all $1\le i\le l$. For the final result we can recover the real integers behind by applying the Chinese remainder theorem. For 
our example of the IMF we choose $l=3$, $p_1=2^{63}-25$, $p_2=2^{63}-165$, and $p_3=2^{63}-259$.\footnote{Choosing primes of the form 
$2^{63}-x$ for small $x$, has the advantage that the computations can be performed using the standard, simple data type \texttt{unsigned long} 
in \texttt{C++}. Our choices are indeed the largest possibilities, see e.g.\ \url{https://primes.utm.edu/lists/2small/0bit.html}. We remark that 
a na\"{\i}ve checking of the primality of the $p_i$ was performed in $41$~seconds. We implement $a=b+c\,\operatorname{mod}\,p$ as $a=b+c$ and 
$\operatorname{if}\,a\ge p\,\operatorname{then}\,a-=p$.}  

The number of losing coalitions is given by $\sum_{x=0}^{q-1} c(x)$ and the number of winning coalitions is given by 
$\sum_{x=q}^{C} c(x)$. Since the total number of coalitions is $2^n$, both numbers can be determined in $O(n\Delta)$ time and $O(\Delta+n)$ space.

For the computation of the Banzhaf index we need to know either the number $c^w(x)$ of coalitions with weight sum $x$ that contain player~$i$ 
or the number $c^{wo}(x)$ of coalitions with weight sum $x$ that do not contain player $i$. For a fixed player $i$ we set 
$c^{wo}(x)=0$ for $0\le x<w_i$. By looping from $w_i$ to $q-1$ we can recursively compute $c^{wo}(x)=c(x)-c^{wo}(x-w_i)$, so that
$\operatorname{Bz}_i^a (v)=\frac{1}{2^{n-1}} \sum_{x=q-w_i}^{q-1} c^{wo}(x)$. Alternatively, we set $c^w(x)=c(x)$ for all $C-w_i<x\le C$ and 
recursively compute $c^w(x)=c(x)-c^w(x+w_i)$ by looping from $C-w_i$ to $q$, so that $\operatorname{Bz}_i^a (v)=\frac{1}{2^{n-1}} 
\sum_{x=q}^{q+w_i-1} c^{w}(x)$.
\begin{Theorem}
  The number of winning, losing coalitions and the Banzhaf indices of all players of a weighted game $v$ can be 
  computed in $O(n\Delta)$ time and $O(\Delta+n)$ space.
\end{Theorem}

\subsection{Counting coalitions per weight and size}
\label{subsec_coalitions_per_weight_and_size}
By $c(x,s)$ we denote the number of coalitions of weight $x$ and cardinality $s$ (for a given weighted game $v$). 
Algorithm~\ref{alg:coalitions_per_weight_forward} and Algorithm~\ref{alg:coalitions_per_weight_backward} can be 
easily adopted to this end. The running time and the memory requirements both increase by a factor of $n$, since 
$0\le s\le n$. We remark $c(x,s)=0$ for $x>\sum_{j=1}^s w_j$ or $x<\sum_{j=n-s+1}^n w_j$, assuming 
$w_1\ge \dots\ge w_n$.\footnote{The players can be sorted in $O(n+\Delta)$ time and space in a preprocessing step.} 
These known values can be taken into account in the boundaries of the for-loops to save time and memory.  By extending the 
definition and recursion for $c^{wo}(x)$, $c^{w}(x)$ to $c^{wo}(x,s)$, $c^{w}(x,s)$, we can state
$$
  \operatorname{SSI}_i(v)=\sum_{s=0}^{n-1} s!(n-s-1)!\cdot \!\!\!\sum_{x=q-w_i}^{q-1} c^{wo}(x,s)
\text{ and }
  \operatorname{SSI}_i(v)=\sum_{s=0}^{n-1} s!(n-s-1)!\cdot \!\!\!\sum_{x=q}^{q+w_i-1} c^{w}(x,s+1).
$$
Of course we can precompute the factorials and the product of the $n-1$ pairs of factorials. 
In our fixed-precision arithmetic we first compute the sums over the $c^{wo}$ or $c^w$ and then 
switch to arbitrary-precision arithmetic.\footnote{We remark $s!(n-s-1)!< 2^{n\log_2 n}$ for $n>1$.} 

\begin{Theorem}
  The $\operatorname{SSI}$ indices of all players of a weighted game $v$ can be 
  computed in $O(n^2\Delta)$ time and $O(n\Delta)$ space.
\end{Theorem}

\subsection{Intersections of weighted games}
Some real-world voting systems are expressed as the intersection of, say $k$, weighted voting games $v_1,\dots,v_k$, i.e., 
a coalition is winning if and only if it is winning in all sub-games $v_1,\dots,v_k$. Let $C_1,\dots,C_k$ be the weights sums 
and $q_1,\dots,q_k$ be the quotas of the sub-games. By easily extending our counting functions $c(x)$ and $c(x,s)$ to 
$c(x_1,\dots,x_k)$ and $c(x_1,\dots,x_k,s)$ we can go along the same lines as in the previous two subsections and 
obtain algorithms with the same complexity bounds if we formally set $\Delta=\min\left\{ \prod_{i=1}^k q_i,\prod_{i=1}^k C_i-q_i+1\right\}$. 
This number may grow very quickly even for moderate values of $k$, so that it may be crucial to choose a representation with a small 
number $k$ of sub-games. We remark that the smallest possible integer $k$ (for a simple game) is called \emph{dimension}. 

\subsection{Computational results}
\label{subsec_computational_results}
We have applied the described algorithms for the four weighted voting games arising from the two different sets of voting weights 
of the IMF in 2015 and 2016, see tables~\ref{table_voting_weights_part1}-\ref{table_voting_weights_part4}, and quotas of either 
85\% or 50\% of the respective weight sums.\footnote{According to the \textit{type} of the decision different values for $q$ are 
used, see e.g.~\cite{leech2013new}. }%For 2015 we have $q=2142486$ or $q=1260286$. For 2016 we have $q=3036989$ or $q=1786465$.} 
All computations were performed on an Intel(R) Core(TM) i7-3720QM cpu with a clock speed of 2.60~GHz and 8~GB RAM. As a a general-purpose 
programming language we have chosen C++ and used the CLN-library\footnote{CLN - Class Library for Numbers, available at 
\url{http://www.ginac.de/CLN}.} for the arbitrary-precision arithmetic parts. 

For 2016 and super-majority, i.e., $q=85\%$, Algorithm 1 needed 4.73 seconds and Algorithm 2 needed 0.67 seconds. The acceleration factor 
for using the described tailored fixed-precision arithmetic over an arbitrary-precision arithmetic is slightly larger than $6$. Using 
pointers instead of the STL class \texttt{vector} results in a speed-up of roughly $2$. The number of winning coalitions 
is given by 4506727722110247822679513808100007271801182981184082. The entire Banzhaf computation, based on Algorithm 2, 
for all players was performed in less than 3~seconds. The corresponding SSI computation took less than 7~minutes.   

We have listed the power distributions for the years 2015 and 2016, the cases of super-majority and simple majority, the power indices 
$\operatorname{Bz}$ and $\operatorname{SSI}$ in tables~\ref{table_voting_power_part1}-\ref{table_voting_power_part4} using a precision of 
five decimal digits for the output. The power of a few countries seem to coincide, which is a numerical artefact, except for France ($i=58$) 
and the United Kingdom ($i=179$). To be more precise, those two countries have the same weights in both 2015 and 2016, so that they are equivalent 
for all values of the quota $q$. For super-majority and simple majority all other countries are inequivalent, which may be seen at the 
exact values of either the Banzhaf or the Shapley-Shubik index. We have written out the exact integers $\operatorname{Bz}_i^a$ for the super-majority 
case in tables~\ref{table_exact_bz_85_part1}-\ref{table_exact_bz_85_part4}. As predicted  by theory, all values have the same parity. For 
the exact values of $n!\cdot \operatorname{SSI}_i$ we 
have
\begin{eqnarray*}
  &\!\!\!& 7943491105336407342763365681397386123932473437094983074178260503548102866231696606 \\
  &\!\!\!& 6280159129991609891845042632480167905404863728522082305141284091873293991567044167 \\
  &\!\!\!& 8913841818287049871073575292808579836984419649634709080444217051613536750323984357 \\
  &\!\!\!& 7219072966973233208480541488235675648000000000000000000000000000000000000000000
\end{eqnarray*}
as an example for the United States in 2015 in the super-majority case. The remaining exact values can 
be obtained from the author upon request.   

Having a closer look at the different power distributions we observe that the choice of the power index or the quota as well as the 
modified weights have a significant impact. The dominance of the United states has further increased from 2015 to 2016. Interestingly enough, 
the Banzhaf power in the super-majority cases shows almost no difference to, e.g., Japan, which is different for the Shapley-Shubik index. 
To obtain a more complete, but still compact, overview about the differences we have introduced 
$\Delta P_y^q(I):=\sum_{i\in I} \left|\operatorname{Bz}_i(v_y^q)-\operatorname{SSI}_i(v_y^q)\right|$, 
$\Delta \operatorname{Bz}^q(I):=\sum_{i\in I} \left|\operatorname{Bz}_i(v_{2015}^q)-\operatorname{Bz}_i(v_{2016}^q)\right|$, and 
$\Delta \operatorname{SSI}^q(I):=\sum_{i\in I} \left|\operatorname{SSI}_i(v_{2015}^q)-\operatorname{SSI}_i(v_{2016}^q)\right|$, 
where $q\in \{85\%,50\%\}$, $y\in\{2015,2016\}$, and $v_y^q$ denotes the corresponding weighted game of the IMF. We evaluate 
those values on the entire set of countries $N$ and on all countries except the biggest five (United States, Japan, Germany, France, 
United Kingdom) $\underline{N}$, see Table~\ref{table_differences}.  

\begin{table}[htp!]
  \caption{Differences in power between years and power indices}{
  \begin{tabular}{rrrrrrrrr}
\hline
  set of countries $I$ & $\Delta P_{2015}^{85\%}$ & $\Delta P_{2016}^{85\%}$ & $\Delta \operatorname{Bz}^{85\%}$ & $\Delta \operatorname{SSI}^{85\%}$ 
  & $\Delta P_{2015}^{50\%}$ & $\Delta P_{2016}^{50\%}$ & $\Delta \operatorname{Bz}^{50\%}$ & $\Delta \operatorname{SSI}^{50\%}$ \\
\hline
$N$  & 54.33\% &         51.98\% &         19.90\% &         27.00\% &         9.46\% &         32.21\% &         46.52\% &       34.68\% \\
$\underline{N}$ & 28.48\% &         26.17\% &         18.45\% &         17.67\% &         2.68\% &         9.70\% &         18.35\% &        18.00\% \\
\hline
  \end{tabular}}
  \label{table_differences}
\end{table}

The stated running times can be easily extrapolated to other examples. If we assume a hypothetical IMF consisting of 1000 members whose weights 
are of similar magnitude as in the 2016 example, then both $n$ and $\Delta$ increase by a factor of $1000/188$. Instead of three primes we would need 
16 primes, so that the computation of the Banzhaf indices in the super-majority case would took approximately 8 minutes while 
the SSI indices may be computed in 4 days.\footnote{In the latter case the memory requirements might become a serious issue if the computations 
for the different primes are not performed consecutively. Computing the intermediate results for each prime in parallel is the better 
option anyway. Using 16 computers (or cores) the mentioned times reduce to seconds and less than 7 hours, respectively.}

\begin{figure}[htp!]
\begin{center}
  \caption{Banzhaf power distribution of the IMF in 2015 for the five largest countries with variable quota}
  \label{fig_bz_2015_q}
  \includegraphics[width=12cm]{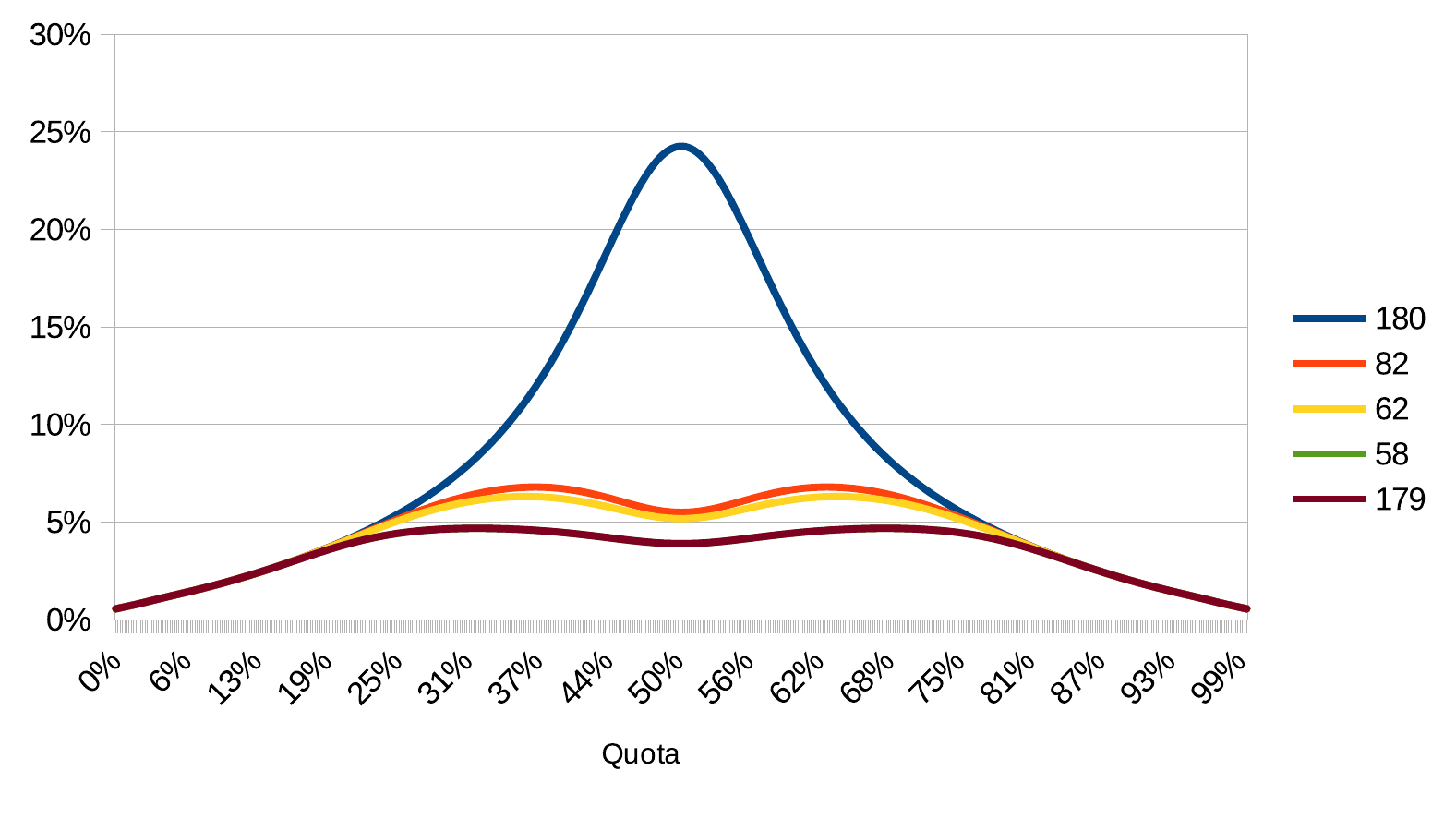}
\end{center}
\end{figure}

\begin{figure}[htp!]
\begin{center}
  \caption{Banzhaf power distribution of the IMF in 2016 for the five largest countries with variable quota}
  \label{fig_bz_2016_q} 
  \includegraphics[width=12cm]{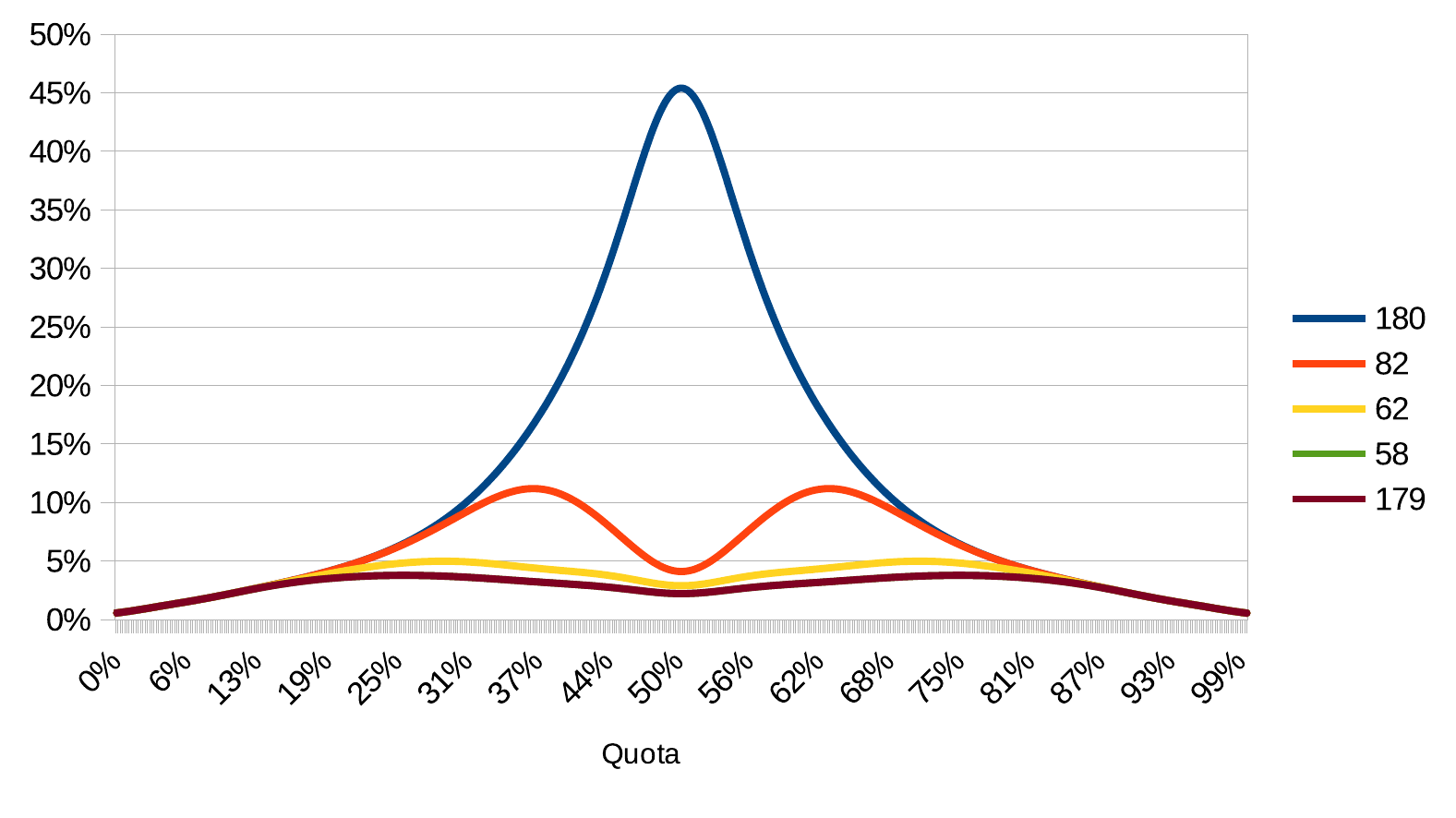}
\end{center}
\end{figure}

The minuscule running time for the Banzhaf index obviously allows more sophisticated applications where the power index computation is performed 
several times. Trying to heuristically solve the inverse power index problem, where weights need to be found whose power distribution is close 
to a given target distribution, is just an example. Here we have computed the Banzhaf power distribution of the IMF when the quota changes 
from 0\% to 100\% in steps of 0.1\%, i.e., 1001 evaluations have been performed, as another possible application. In 
figures~\ref{fig_bz_2015_q}-\ref{fig_bz_2016_q} we have depicted the corresponding power for the five most powerful countries 
(United States $i=180$, Japan $i=82$, Germany $i=62$, France $i=58$, United Kingdom $i=179$). We can see that the respective Banzhaf indices 
are rather volatile with respect to changes of the quota. The difference between the respective Banzhaf power shares is negligible for 
Japan and Germany, while there is no difference between France and the United Kingdom, for all values of the quota $q$. For an extreme 
quota of 0\% or 100\% all countries obtain exactly the same Banzhaf power share. For quotas 
below 15\% or above 85\% there is almost no difference in power for the five largest countries. However, there is a critical interval, 
say $q$ between 25\% and 75\%, where the relative power distribution between the five largest countries is very sensitive to changes of the quota. 
The United States most intensive benefit from quotas around 50\%. In 2016 also the Banzhaf power share of Japan is very sensitive to changes 
of the quota. Instead of 50\% a quota of roughly 65\% would be rather favorable for them.      

\section{Conclusions}
\label{sec_conclusion}
Nevertheless the computation of both the Banzhaf and the Shapley-Shubik index is NP-hard for weighted voting games, we have 
demonstrated that in practice it is not too hard to compute the exact values if the considered games are not \textit{too large}. 
In the used sense, the current IMF voting system is definitely not too large since the Banzhaf indices can be computed in seconds 
and the Shapley-Shubik indices can be computed in a few minutes. For weighted games of that magnitude no approximations to the 
real values are necessary.

Even more, an efficient computation does not rely on sophisticated algorithms but low-level details in order to gain speed-up factors. 
For $C\ll 2^n$, which should be the case for all non-tiny real--world examples, the use of generating function approaches yields 
no benefit, although being a common topic in the literature. The used underlying idea of counting coalitions per weight and size by 
a simple recursion was just enhanced  by allowing the reverse direction starting from the weight sum $C$ for quotas larger than 50\%. 
There is a single small insight that allows to recover those counts for the cases where a certain player is either assumed to be part 
or not to be part of the counted coalitions more efficiently than a direct enumeration. Using this approach the complexity for 
computing the considered indices are (up to a small constant) the same for a single player and all players. 

For our real--world example of the IMF, the resulting power distributions are rather different from the weight shares and between 
diverse power indices like the Banzhaf and the Shapley-Shubik index. We suspect that this is not a numerical artefact of this specific 
example, so that it might be a good idea to compute several power indices to get a more comprehensive view whenever the considered 
committee has some non-negligible impact.  

The distribution of the Banzhaf power shares is rather sensitive to changes of the quota and there are clear incentives 
for the few largest countries to alter them in their sense. The conclusion that may be drawn from that fact is debatable and the 
choice of the quota should indeed obtain more consideration. 

% Bibliography
%\bibliographystyle{plain}
%\bibliography{imf}

\begin{appendix}{Tables of voting weights and the power distribution}
\begin{table}[htp!]
  \caption{Voting weights in the IMF -- part 1}{
\begin{tabular}{rrrrrr}
\hline
index & member/year & 2015	& &			2016			\\
\hline
 1 & Afghanistan, Islamic Republic of	&	2357	&	0.094\%	&	2665	&	0.075\%	\\
 2 & Albania	&	1338	&	0.053\%	&	2439	&	0.068\%	\\
 3 & Algeria	&	13285	&	0.527\%	&	13593	&	0.380\%	\\
 4 & Angola	&	3601	&	0.143\%	&	3909	&	0.109\%	\\
 5 & Antigua and Barbuda	&	873	&	0.035\%	&	1246	&	0.035\%	\\
 6 & Argentina	&	21909	&	0.869\%	&	22217	&	0.622\%	\\
 7 & Armenia	&	1658	&	0.066\%	&	1966	&	0.055\%	\\
\end{tabular}}
  \label{table_voting_weights_part1}
\end{table}

\begin{table}[htp!]
  \caption{Voting weights in the IMF -- part 2}{
  \begin{tabular}{rrrrrr}
\hline
index & member/year & 2015	& &			2016			\\
\hline
 47 & Dominica 	&	820	&	0.033\%	&	1161	&	0.032\%	\\
 48 & Dominican Republic	&	2927	&	0.116\%	&	3235	&	0.091\%	\\
 49 & Ecuador	&	4216	&	0.167\%	&	4524	&	0.127\%	\\
 50 & Egypt	&	10175	&	0.404\%	&	10483	&	0.293\%	\\
 51 & El Salvador	&	2451	&	0.097\%	&	2759	&	0.077\%	\\
 52 & Equatorial Guinea	&	1261	&	0.050\%	&	1569	&	0.044\%	\\
 53 & Eritrea	&	897	&	0.036\%	&	1205	&	0.034\%	\\
\end{tabular}}
  \label{table_voting_weights_part2}
\end{table}

\begin{table}[htp!]
  \caption{Voting weights in the IMF -- part 3}{
  \begin{tabular}{rrrrrr}
\hline
index & member/year & 2015	& &			2016			\\
\hline
 95 & Liberia	&	2030	&	0.081\%	&	2338	&	0.065\%	\\
 96 & Libya	&	11975	&	0.475\%	&	12283	&	0.344\%	\\
 97 & Lithuania	&	2577	&	0.102\%	&	5462	&	0.153\%	\\
 98 & Luxembourg	&	4925	&	0.195\%	&	5233	&	0.146\%	\\
 99 & Macedonia, former Yugoslav Republic of	&	1427	&	0.057\%	&	2449	&	0.069\%	\\
100 & Madagascar	&	1960	&	0.078\%	&	2268	&	0.063\%	\\
101 & Malawi	&	1432	&	0.057\%	&	2434	&	0.068\%	\\
\hline
  \end{tabular}}
  \label{table_voting_weights_part3}
\end{table}

\begin{table}[htp!]
  \caption{Voting weights in the IMF -- part 4}{
  \begin{tabular}{rrrrrr}
\hline
index & member/year & 2015	& &			2016			\\
\hline
142 & Saudi Arabia	&	70593	&	2.801\%	&	70901	&	1.984\%	\\
143 & Senegal	&	2356	&	0.093\%	&	2664	&	0.075\%	\\
144 & Serbia	&	5415	&	0.215\%	&	7594	&	0.213\%	\\
145 & Seychelles	&	847	&	0.034\%	&	1275	&	0.036\%	\\
146 & Sierra Leone	&	1775	&	0.070\%	&	2083	&	0.058\%	\\
147 & Singapore	&	14818	&	0.588\%	&	15126	&	0.423\%	\\
148 & Slovak Republic	&	5013	&	0.199\%	&	5321	&	0.149\%	\\
\hline
  \end{tabular}}
  \label{table_voting_weights_part4}
\end{table}

\begin{table}[htp!]
  \caption{Voting power in the IMF -- part 1}{
  \begin{tabular}{rrrrrrrrr}
\hline
    & \multicolumn{4}{c}{supermajority} & \multicolumn{4}{c}{simple majority} \\
    & \multicolumn{2}{c}{$\operatorname{Bz}_i$} & \multicolumn{2}{c}{$\operatorname{SSI}_i$} & \multicolumn{2}{c}{$\operatorname{Bz}_i$} & \multicolumn{2}{c}{$\operatorname{SSI}_i$} \\   
$i$ & 2015 & 2016	&  2015 & 2016 & 2015 & 2016	&  2015 & 2016 \\
\hline
1        &        0.153\%        &        0.126\%        &        0.088\%        &        0.074\%        &        0.086\%        &        0.056\%        &        0.088\%        &        0.067\%        \\
2        &        0.087\%        &        0.115\%        &        0.050\%        &        0.068\%        &        0.049\%        &        0.051\%        &        0.050\%        &        0.062\%        \\
3        &        0.845\%        &        0.634\%        &        0.497\%        &        0.378\%        &        0.483\%        &        0.286\%        &        0.501\%        &        0.345\%        \\
4        &        0.234\%        &        0.184\%        &        0.134\%        &        0.108\%        &        0.131\%        &        0.082\%        &        0.135\%        &        0.099\%        \\
5        &        0.057\%        &        0.059\%        &        0.032\%        &        0.034\%        &        0.032\%        &        0.026\%        &        0.033\%        &        0.032\%        \\
\hline
  \end{tabular}}
  \label{table_voting_power_part1}
\end{table}

\begin{table}[htp!]
  \caption{Voting power in the IMF -- part 2}{
  \begin{tabular}{rrrrrrrrr}
\hline
    & \multicolumn{4}{c}{supermajority} & \multicolumn{4}{c}{simple majority} \\
    & \multicolumn{2}{c}{$\operatorname{Bz}_i$} & \multicolumn{2}{c}{$\operatorname{SSI}_i$} & \multicolumn{2}{c}{$\operatorname{Bz}_i$} & \multicolumn{2}{c}{$\operatorname{SSI}_i$} \\   
$i$ & 2015 & 2016	&  2015 & 2016 & 2015 & 2016	&  2015 & 2016 \\
\hline
48        &        0.190\%        &        0.152\%        &        0.109\%        &        0.090\%        &        0.106\%        &        0.068\%        &        0.110\%        &        0.082\%        \\
49        &        0.274\%        &        0.213\%        &        0.157\%        &        0.125\%        &        0.153\%        &        0.095\%        &        0.158\%        &        0.115\%        \\
50        &        0.653\%        &        0.491\%        &        0.380\%        &        0.291\%        &        0.370\%        &        0.221\%        &        0.383\%        &        0.266\%        \\
51        &        0.159\%        &        0.130\%        &        0.091\%        &        0.076\%        &        0.089\%        &        0.058\%        &        0.092\%        &        0.070\%        \\
52        &        0.082\%        &        0.074\%        &        0.047\%        &        0.043\%        &        0.046\%        &        0.033\%        &        0.047\%        &        0.040\%        \\
\hline
  \end{tabular}}
  \label{table_voting_power_part2}
\end{table}

\begin{table}[htp!]
  \caption{Voting power in the IMF -- part 3}{
  \begin{tabular}{rrrrrrrrr}
\hline
    & \multicolumn{4}{c}{supermajority} & \multicolumn{4}{c}{simple majority} \\
    & \multicolumn{2}{c}{$\operatorname{Bz}_i$} & \multicolumn{2}{c}{$\operatorname{SSI}_i$} & \multicolumn{2}{c}{$\operatorname{Bz}_i$} & \multicolumn{2}{c}{$\operatorname{SSI}_i$} \\   
$i$ & 2015 & 2016	&  2015 & 2016 & 2015 & 2016	&  2015 & 2016 \\
\hline
95        &        0.132\%        &        0.110\%        &        0.075\%        &        0.065\%        &        0.074\%        &        0.049\%        &        0.076\%        &        0.059\%        \\
96        &        0.765\%        &        0.574\%        &        0.448\%        &        0.342\%        &        0.435\%        &        0.258\%        &        0.451\%        &        0.312\%        \\
97        &        0.167\%        &        0.257\%        &        0.096\%        &        0.151\%        &        0.094\%        &        0.115\%        &        0.097\%        &        0.138\%        \\
98        &        0.319\%        &        0.246\%        &        0.183\%        &        0.145\%        &        0.179\%        &        0.110\%        &        0.185\%        &        0.133\%        \\
99        &        0.093\%        &        0.115\%        &        0.053\%        &        0.068\%        &        0.052\%        &        0.052\%        &        0.054\%        &        0.062\%        \\
\hline
  \end{tabular}}
  \label{table_voting_power_part3}
\end{table}

\begin{table}[htp!]
  \caption{Voting power in the IMF -- part 4}{
  \begin{tabular}{rrrrrrrrr}
\hline
    & \multicolumn{4}{c}{supermajority} & \multicolumn{4}{c}{simple majority} \\
    & \multicolumn{2}{c}{$\operatorname{Bz}_i$} & \multicolumn{2}{c}{$\operatorname{SSI}_i$} & \multicolumn{2}{c}{$\operatorname{Bz}_i$} & \multicolumn{2}{c}{$\operatorname{SSI}_i$} \\   
$i$ & 2015 & 2016	&  2015 & 2016 & 2015 & 2016	&  2015 & 2016 \\
\hline
142        &        2.725\%        &        2.609\%        &        2.742\%        &        2.028\%        &        2.556\%        &        1.474\%        &        2.717\%        &        1.821\%        \\
143        &        0.153\%        &        0.126\%        &        0.088\%        &        0.074\%        &        0.086\%        &        0.056\%        &        0.088\%        &        0.067\%        \\
144        &        0.351\%        &        0.357\%        &        0.202\%        &        0.211\%        &        0.197\%        &        0.160\%        &        0.204\%        &        0.192\%        \\
145        &        0.055\%        &        0.060\%        &        0.031\%        &        0.035\%        &        0.031\%        &        0.027\%        &        0.032\%        &        0.032\%        \\
146        &        0.115\%        &        0.098\%        &        0.066\%        &        0.058\%        &        0.065\%        &        0.044\%        &        0.067\%        &        0.053\%        \\
\hline
  \end{tabular}}
  \label{table_voting_power_part4}
\end{table}

\begin{table}[htp!]
  \caption{Exact Banzhaf counts $\mathbf{\operatorname{Bz}_i^a}$ for supermajority -- part 1}{
  \begin{tabular}{rrr}
\hline
$i$ & 2015 & 2016	\\
\hline 
\tiny{1} & \tiny{49492696550469334288388474302188300702732503278942} & \tiny{176903926397724458899541247938651532746047517818906} \\
\tiny{2} & \tiny{28108818684726200071728584276680195673408532632506} & \tiny{161911654209258251928883011914306328209189949330986} \\
\tiny{3} & \tiny{273067686459011964864016012688469258989855194791048} & \tiny{894022914523707629419913395215803955826725915570768} \\
\tiny{4} & \tiny{75544231177486375978776859223371297251582054333762} & \tiny{259370900172430505766454804559359034798947935884810} \\
\tiny{5} & \tiny{18342422447942989488068768527179198893754345719314} & \tiny{82733927205579511449928590811181587395707525770476} \\
\tiny{6} & \tiny{433741858747716854195909526475817535173470581906728} & \tiny{1438072991668218360540956801914813269126044117351308} \\
\tiny{7} & \tiny{34827220802205252181870565955690262877158934233490} & \tiny{130525954991744089345193787551781828973012811299238} \\
\hline
  \end{tabular}}
  \label{table_exact_bz_85_part1}
\end{table}

\begin{table}[htp!]
  \caption{Exact Banzhaf counts $\mathbf{\operatorname{Bz}_i^a}$ for supermajority -- part 2}{
  \begin{tabular}{rrr}
\hline
$i$ & 2015 & 2016	\\
\hline 
\tiny{48} & \tiny{61438462001559665113999981127215796585984210855648} & \tiny{214703239153916461314845491769597161273978081883676} \\
\tiny{49} & \tiny{88392867735151915207586461901703825857712334268926} & \tiny{300096595204460456827471144043187771118052667702764} \\
\tiny{50} & \tiny{211053904455260606918250786934031134685520223154414} & \tiny{692159614227603811889517072645010455131000993604518} \\
\tiny{51} & \tiny{51463609924728845688958697853656613736623382622284} & \tiny{183138834038284691902924536354058860926546225596384} \\
\tiny{52} & \tiny{26491860599355877436888623101254938092665600267626} & \tiny{104176078522686009873370603749878533351537159776114} \\
\tiny{53} & \tiny{18846581130575530182651709645454563684490631687646} & \tiny{80011960911102504338320129319511747538362140926976} \\
\tiny{54} & \tiny{35226046701012768753027566935295199597069035091906} & \tiny{231076410124979151881162099161280586981496993085196} \\
\hline
  \end{tabular}}
  \label{table_exact_bz_85_part2}
\end{table}

\begin{table}[htp!]
  \caption{Exact Banzhaf counts $\mathbf{\operatorname{Bz}_i^a}$ for supermajority -- part 3}{
  \begin{tabular}{rrr}
\hline
$i$ & 2015 & 2016	\\
\hline 
\tiny{95} & \tiny{42633958042491066739671348871443478321386940033530} & \tiny{155210719725738127825182299895255583198461342494118} \\
\tiny{96} & \tiny{247157849931994416381682310743110238430008662255444} & \tiny{809284867620330781092103774503572607880975700634310} \\
\tiny{97} & \tiny{54104934213516959447392068269135858247168076667212} & \tiny{362141825341601765105298798708663702946011014384032} \\
\tiny{98} & \tiny{103174011727540055347125360906146890295086611388140} & \tiny{347002831295754786039688225426685425553829113352468} \\
\tiny{99} & \tiny{29977613645414705823501642390509810402839894970228} & \tiny{162575085386940526359782345079435496274014948315940} \\
\tiny{100} & \tiny{41165262535925141133298445653237234427829732414728} & \tiny{150566216658081937850639757955081825600451573581624} \\
\tiny{101} & \tiny{30082596933358311863367988305279579489729633493986} & \tiny{161579936725190284284386840690183433473870373636974} \\
\hline
  \end{tabular}}
  \label{table_exact_bz_85_part3}
\end{table}

\begin{table}[htp!]
  \caption{Exact Banzhaf counts $\mathbf{\operatorname{Bz}_i^a}$ for supermajority -- part 4}{
  \begin{tabular}{rrr}
\hline
$i$ & 2015 & 2016	\\
\hline 
\tiny{142} & \tiny{880719880889746205050554576253671936253306729604848} & \tiny{3676265198068175422113349665153212813969097013199656} \\
\tiny{143} & \tiny{49471727597057335306131923257174325184458483338542} & \tiny{176837594941662819439217298459469758549843454999802} \\
\tiny{144} & \tiny{113367090641059940461002376473927867025289026259444} & \tiny{502769941764486824601335626776974734248420124375276} \\
\tiny{145} & \tiny{17796241582555972529647845427695103090190196272002} & \tiny{84659194321239810261578274676558081484774259150292} \\
\tiny{146} & \tiny{37282994499005042799257701148151378617750538909674} & \tiny{138290369392939768640092980732285176104943285779542} \\
\tiny{147} & \tiny{302950294620727927003732396031052197317326829457440} & \tiny{992581627260105769126385986812033471596368636275590} \\
\tiny{148} & \tiny{105006025900540763870469317668724012845234490381094} & \tiny{352821120109159596707036375327612661869380642957788} \\
\hline
  \end{tabular}}
  \label{table_exact_bz_85_part4}
\end{table}

\end{appendix}

\end{document}